\def\12{{1\over2}}
\def\bi{\bigskip}
\def\noi{\noindent}
\def\be{\begin{equation}}
\def\en{\end{equation}}
\def\bq{\begin{eqnarray}}
\def\eq{\end{eqnarray}}
\def\bc{\begin{center}}
\def\ec{\end{center}}
\def\beit{\begin{itemize}}
\def\eit{\end{itemize}}

\documentstyle[prl,aps,epsf]{revtex}

\begin{document}
\draft
\title{$\phi\to\pi^0\pi^0\gamma$ DECAY WITHIN A $U(3)\times U(3)$ LINEAR SIGMA 
MODEL}

\vskip2ex

\author{ J. L. Lucio \\
{\em Instituto de F\'{\i}sica, Universidad de 
Guanajuato}\\ 
{\em Loma del Bosque \# 103, Lomas del Campestre, 
37150 Le\'on, Gto.; M\'exico.} \\
and\\
{\em Laboratori Nazionali di Frascati, P.O. Box 13, I-00044, Frascati,Italy.}}
 
\author{ M. Napsuciale \\ 
{\em Instituto de F\'{\i}sica, Universidad de Guanajuato}\\ 
{\em Loma del Bosque \# 103, Lomas del Campestre, 
37150 Le\'on, Gto.; M\'exico } }

\maketitle
\baselineskip=11.6pt
\begin{abstract}
We show that the recently observed pion invariant mass distribution of the
 $\phi\to\pi^0\pi^0\gamma$ decay can be satisfactorily described by the
chiral $U(3)\times U(3)$ Linear Sigma Model.
\end{abstract}
\baselineskip=14pt
\section{Introduction}
Over  the past years 
experimental evidence has accumulated for the existence of  light scalar 
mesons \cite{whs}, and different proposals exist for the $\bar qq$  lowest 
lying scalar meson nonet. The Particle Data Group (PDG) \cite{pdg} 
candidates for the 
ground state $\bar q q$ scalar  nonet are : the $f_0(980)$, $f_0(1370)$ and 
the  $f_0(400-1250)$ ( or $\sigma$ meson)  for two 
states in 
the  $I=0$ sector; the $a_0(980)$ and $a_0(1450)$ for the isovector 
scalar meson, and the 
$K^*_0(1430)$ for the isospinor scalar meson.

The small decay rate into two photons is among the most
important drawbacks for 
the identification of the $a_0(980)$ and $f_0(980)$ as the $\bar q q$ scalar 
isovector and isosinglet respectively.
These decay rates have been calculated using a  
variety of approaches \cite{qmod,molecule,fourq}, in particular, in  
different versions 
of the quark model \cite{qmod}. The generally accepted conclusion, 
is that the $a_0(980), f_0(980) \to\gamma\gamma$ decay widths are not 
consistent with a $q\bar q$ structure.  

Moreover, the nearby mass degeneracy of these mesons suggest they are the 
scalar analogous of the  $\omega$ and $\rho$ system, i.e. the  
$a_0(980), f_0(980$ 
are expected to be  $\bar qq$ ( $q=u,d$ ) which however contradicts the strong 
coupling of the $f_0(980)$ meson to the $\bar KK$ system.  

Other possibilities such as a molecule picture \cite{molecule} 
and a $\bar q q \bar q q $ structure \cite{fourq} have been explored. 
Recently, it has been argued that the four-quark picture for these mesons 
is consistent not only with the two photon decays of these
states but also with the $\Phi\to f_0\gamma\to\pi\pi\gamma$ decay 
\cite{achasov}.

An alternative approach to the hadron physics in this energy region is
provided by the $U(3)\times U(3)$ chiral model which incorporates a nonet of
scalar as well as a nonet of pseudoscalar 
particles \cite{gasiorowicz,schechter,scadron,napsu,thooft,tornq}.  In fact, 
the $U_A(1)$ component 
of the 
$U(3)\times U(3)$ symmetry exhibited by the light sector of QCD in the 
massless 
quark limit,  is broken at the quantum level which in the model amounts to the
possibility of incorporating otherwise forbidden terms in the interaction
lagrangian. In this model, the
$f_0(980)$ turns out to be a mostly 
$\bar s s$ meson whereas the $a_0(980)$ meson is the chiral partner of the
pion \cite{napsu,tornq}; the reason for the nearby degeneracy of the 
$a_0(980)$ and the $f_0980)$ 
being that the $U_A(1)$ anomaly pushes up the $a_0(980)$ mass while leaving 
untouched the $f_0(980)$ meson.

The model has shown to be phenomenologically succesful  
\cite{napsu,tornq,luna}. In particular 
the $a_0(980)\to \gamma\gamma$ and 
$f_0(980)\to \gamma\gamma$ decays are consistenly accounted for 
in this framework  \cite{luna}, providing thus an explanation to the failure
of the quark model calculations which do not take into account 
$U_A(1)$-breaking induced interactions.

\bi
This year, DA$\phi$NE, the  high luminosity $\phi$ factory, will perform 
precise measuraments of $\phi$ radiative decays. The Novosibirsk CMD and SND
collaborations already reported, among others,  measurements of 
$\phi\to \pi^0\pi^0 (\eta)\gamma$ and 
$\pi^+\pi^-\gamma$ \cite{exp}. On the theoretical side the $\phi \to 
\pi\pi\gamma$ has been considered by a number of authors \cite{achasov,bramon1,bramon2,bramon3,nalu,oset}. In particular Bramon, Grau and 
Pancheri (B.G.P.)
considered vector meson and chiral loop contributions. By itself the vector
meson contribution turns out to be small whereas the chiral loops lead to a
broad pion invariant mass spectrum which could easily be distinguished by 
experiments.

In this contribution we report calculations for the 
$\phi \to \pi^0\pi^0\gamma$
 decay within the $U(3)\times U(3)$ model where intermediate scalar resonances 
naturally appear. From the theoretical point of view this is a clean
process, since no final state radiation exists (as compared to the decay in 
charged pions) and the pseudoscalar mixing angle is not involved 
(as in the $\pi^0\eta\gamma$ case).

\section{Scalar meson contributions to $\phi\to \pi^0\pi^0\gamma$}

The process under consideration is generated at one loop level. The diagrams 
contributing to this process are depicted in Fig1.

\begin{figure}[t]
\vspace{9.0cm}
\includegraphics{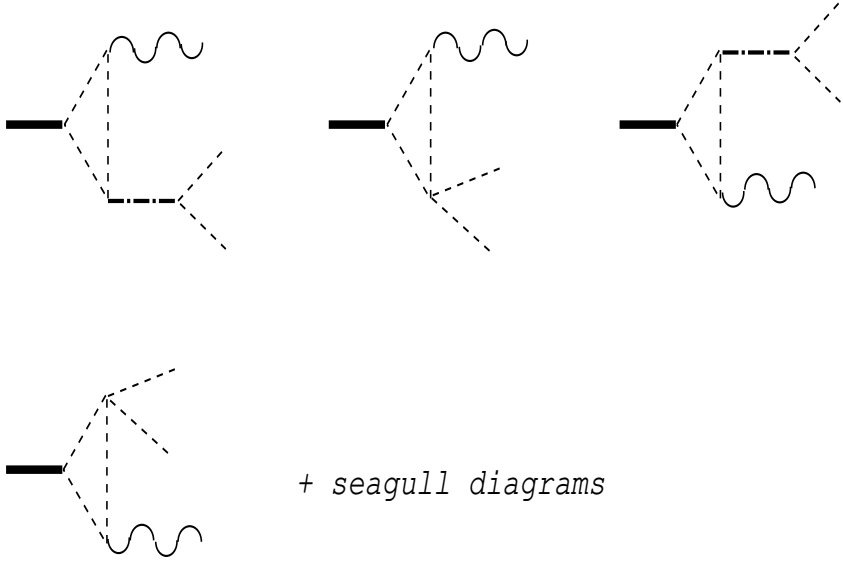}
 \caption{\it
      Contributions to $\phi\to\pi^0\pi^0\gamma$ in the LSM. Dashed lines 
denote pseudoscalar mesons (kaons in the loops and neutral pions in the final 
state) while dot-dashed lines denote (intermediate) isoscalar scalar mesons 
(sigma and $f_0(980)$). 
    \label{fig1} }
\end{figure}

\noi The amplitude arising from the scalar contributions is given by:

\be
{\it M}(\phi\to\pi^0\pi^0\gamma)= e~G(m^2_{\pi\pi})~T_{\mu\nu}
\eta^\mu\epsilon^\nu
\en

\noi where
\be
T_{\mu\nu}=Q.k~g_{\mu\nu}-k_\mu Q_\nu
\en
and
\be
G(m^2_{\pi\pi})= {g_{\phi K^+K^-}~\over 2\pi^2M^2_\phi} F L(m^2_{\pi\pi})
\label{loopcoup}
\en
\noi with the loop function
\be
L(m^2_{\pi\pi})= {1\over 2(a-b)} - {2\over(a-b)^2}~[f({1\over b}) - f({1\over a})]
+{a\over(a-b)^2}~[g({1\over b})-g({1\over a})].   
\en
\noi where

\[ f(z)  = \left\{ \begin{array}{ll}
 - \bigg ( arcsin \big [\frac{1}{2 \sqrt {z}}  \big ] \bigg)^2 &
 \mbox{ $z > \frac{1} {4}$} \\ 
 \frac{1}{4} \bigg ( ln { \eta_+ \over \eta_-} - i \pi \bigg )^2 &
 \mbox{$ z < \frac{1}{4}$}  \end{array} \right. \]

\[ g(z)  = \left\{ \begin{array}{ll}
  (4z - 1)^{1 \over 2} \arcsin \big [ { 1 \over 2 \sqrt{z}} \big ]  &
 \mbox{ $z > \frac{1} {4}$} \\ 
 {1 \over 2} (1 - 4 z )^{1 \over2} ( \ln {\eta_+ \over \eta_-} -i \pi ) &
 \mbox{$ z < \frac{1}{4}$}  \end{array} \right. \]

\noi with

\begin{equation}
\eta_{\pm} = {1 \over 2} [1 \pm ( 1 - 4 z)^{1 \over 2}  ] , ~~~~~
 a ={ M^2_\phi \over 
m^2_{k^+} } , ~~ b = {m^2_{\pi\pi} \over m^2_{k^+}}. 
\end{equation}

The F factor appearing in Eq.(\ref{loopcoup}) contain the information on the 
coupling constants.

\be
F= 2(g_{KK\pi\pi}-{g_{\sigma\pi\pi}g_{\sigma KK}\over m^2_{\pi\pi}-
m^2_\sigma +i\Gamma_\sigma m_\sigma}-{g_{f\pi\pi}g_{fKK}\over m^2_{\pi\pi}-
m^2_f +i\Gamma_f m_f})
\en

The three and four-meson couplings are given by the 
model \cite{schechter,napsu,tornq} as

\bq \nonumber
g_{\sigma KK}&=&- {m^2_\sigma -m^2_K \over 2f_K}(cos\phi- 
                     \sqrt{2} sin\phi)~~~ ;~~~g_{\sigma \pi\pi}=- 
{m^2_\sigma -m^2_\pi \over 2f_\pi}cos\phi \\ \label{couplings}
g_{f KK}&=&- {m^2_f -m^2_K \over 2f_K}(sin\phi+ 
                    \sqrt{2}cos\phi) ~~~;~~~
g_{f\pi\pi}=- {m^2_\sigma -m^2_\pi \over 2f_\pi}sin\phi \\ \nonumber
g_{KK\pi\pi}&=&-{m^2_\sigma-m^2_K\over 4f_Kf_\pi}. \\ \nonumber
\eq

The pion invariant mass spectrum is obtained  

\be
{d\Gamma \over dm_{\pi\pi}}= {\alpha_{em}\over 4\pi}{m_{\pi\pi}\over M_\phi}
({g_{\phi KK}\over 4\pi})^2({1\over 4\pi})^2({M_\phi\over m_K})^4 
|L(m^2_{\pi\pi})|^2|F|^2(1-{m^2_{\pi\pi}\over M^2_\phi})^3\sqrt{1-{4m^2_\pi 
\over m^2_{\pi\pi}}} \label{spectrum}
\en

\begin{figure}[t]
\vspace{9.0cm}
\includegraphics{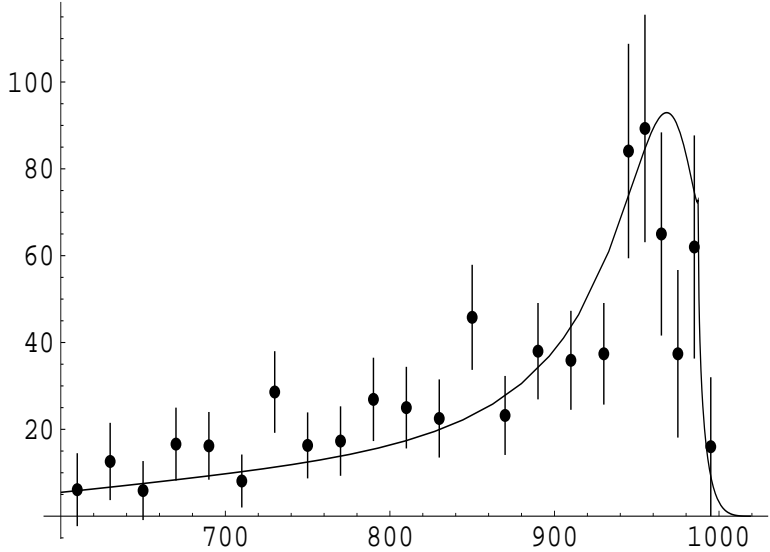}
\caption{\it
 $dB(\phi\to\pi^0\pi^0\gamma)/dm_{\pi\pi} \times 10^{-8} MeV^{-1}$ 
as a function of the dipion invariant mass. The experimental points are taken 
from the SND Coll. V.M. Aulchenko et.al. [14].
\label{Fig.2}  }
 \end{figure}

We observe from Eqs(7,8) that the energy spectrum depend on the scalar
mixing angle (in the $\{ |S>,|NS>\}$ basis) 
$\phi$ and the scalar meson masses. In particular is highly sensitive to 
the chosen value for the mixing angle. It is worth to remark that for a 
sigma meson mass above 600 MeV the theoretical predictions for the energy 
spectrum yield a desastrous result as compared with the experimental 
results. Our results reduces to those of B.G.P. \cite{bramon1} -upto an overall
normalization factor- in the very heavy (and non-mixed) scalars limit (as 
compared to the typical 1GeV scale). We have been unable to trace back the 
difference in normalization of the two approaches (a factor $\sqrt(3)$ in the 
amplitud).

The LSM results for the energy spectrum in Eq.(\ref{spectrum}) are shown in 
Fig. 2. We use $\phi=-9^0$, $m_f=980$ MeV, $\Gamma_f=70$MeV and  
$m_\sigma =560$ MeV in the numerical evaluations. The sigma width is dictated 
by the model as

\be
\Gamma_\sigma= {3m^3_\sigma \over 32 \pi f^2_\pi}((1-
{m^2_\pi \over m^2_\sigma})cos(\phi))^2\sqrt{1-4 {m^2_\pi \over m^2_\sigma}}.
\label{Gsig}
\en
\bi

In table 1 we also show the theoretical predictions arising from Eq(
\ref{spectrum}) for 
the total and partial (i.e. integrated over a limited region of the energy 
spectrum) Branching Ratios. For comparison we also included the experimental 
results reported by the Novosibirsk groups. Within experimental errors, 
agreement is satisfactory.

\bi
\bc
{\bf Table 1}
\ec

\bi

\bc
\begin{tabular}{|c|c|c|c|} \hline
  $m_{\pi^0\pi^0}(MeV)$ & $BR(CMD-2)(\times 10^{-4})$ & 
$BR(SND)(\times 10^{-4})$ & $BR_{TH}(\times 10^{-4})$       \\ \hline
\hline
   $> 550 $    &$ 1.06\pm0.09\pm0.06 $ &         
&$0.99$   \\ \hline
   $>700  $    &$ 0.92\pm0.08\pm0.06 $ & $1.00\pm0.07\pm 0.12 $      
& $0.90 $\\  \hline
   $>900 $     &$0.57\pm0.06\pm0.04  $ & $0.50\pm0.06\pm 0.06 $       
&$0.53  $ \\      \hline
   $total(>2m_\pi)$   &$1.08\pm0.17\pm0.09  $ & $1.14 \pm 0.10\pm 0.12$
 &$1.08 $ \\ \hline
\end{tabular}
\ec
\bi

So far we have used a Breit-Wigner to describe the 
sigma (and $f_0(980)$) propagator. It has been argued that 
 the inclusion of the sigma width in this way strongly breaks chiral 
symmetry \cite{achasov2,lnm}. If 
we modify the sigma vertices in such a way that the  Goldstone Boson nature 
of the pions is preserved as discussed in \cite{lnm}, the curve in Fig. 2  
is modified. In this case agreement with 
experimental results is obtained for a lower sigma mass $m_\sigma =430 MeV$ and
the same mixing angle $\phi=-9^\circ $.

Summarizing, the VEPP-2M SND and CMD2 experimental results for the 
$\phi\to \pi^0\pi^0\gamma$ results are consistent with a mostly 
$\bar ss$ $f_0(980)$ meson provided we take into account the effects of the 
$U_A(1)$ breaking in the scalar sector. This process gives also support to 
the existence of a scalar meson resonance ($\sigma$) in the 400-600 MeV.
The process under consideration is highly sensitive to the scalar mixing 
angle and experimental results for this process require $\phi\approx -9^0$ 
which is consistent with other estimates \cite{napsu}.

\section{Acknowledgements}
This work was supported by Conacyt Mexico under project I27604-E.
One of us (J.L.L.) whish to acknowledge financial support from Conacyt, 
Mexico and Concyteg-Guanajuato, Mexico.

\end{document}